\documentclass{svproc}
\usepackage{url}
\usepackage{amsmath}
\usepackage{graphicx}
\usepackage{bm}

\begin{document}
\mainmatter              
%
\title{Emergence of edge state in suspension of self-propelled particles}
\titlerunning{Edge state in suspension of self-propelled particles}  
%
%
%
%
\author{Yoshiki Hiruta\inst{1} \and Kenta Ishimoto\inst{2}}
\authorrunning{Yoshiki Hiruta \and Kenta Ishimoto} 
%
\tocauthor{}
%

\institute{Department of Physics and Astronomy, Faculty of Science and Technology, Tokyo University of Science, Chiba 278-8510, Japan
 \email{hiruta@rs.tus.ac.jp},
 \and
Department of Mathematics, Kyoto University, Kyoto, Japan\\
\email{kenta.ishimoto@math.kyoto-u.ac.jp}
}

\maketitle              

\begin{abstract}
We numerically study a model convection system of a suspension of self-propelled particles, motivated by recent experimental findings of localized and bistable bioconvection pattern, being distinct from classical Rayleigh--B\'{e}nard convection.
Linear stability analysis of the model system reveals that the trivial noncovection state is stabilized by an increase of self-propelled speed in the vertical direction. 
Through numerical simulations, we found a nonlinear convection state even when the nonconvection state is stable. 
Applying ideas and tools developed in wall-bounded flows,
we numerically identified an edge state, which is 
an unstable solution on a basin boundary in the model dynamical systems.
\end{abstract}
\section{Introduction}
A suspension of swimming microorganisms often exhibits convective flow patterns, 
known as bioconvection~\cite{Childress,Taheri,ref_3}.
In typical settings, bioconvection patterns sustain for a longtime 
if the average density of a suspension exceeds a critical value as seen in the Rayleigh--B\'{e}nard convection.
Recently, Yamashita et al. \cite{ref_3} experimentally manipulated a suspension of microorganisms 
that swim away from the light source at the bottom of a chamber,
and reported an emergence of localized pattern and bistability at the onset of the transition,
which cannot happen in Rayleigh--B\'{e}nard convection \cite{Joseph}.

Localization and bistability are well studied in wall-bounded shear flows~\cite{ref_9,ref_11,ref_8} 
and other flows in a periodic box~\cite{ref_4,ref_5} and 
analytical methods based on dynamical systems theory have been developed.
In particular, the existence of bistable turbulent localized pattern is well characterized by a so-called {\it edge state} \cite{ref_8,Itano}, which is an invariant set on a basin boundary. The phase space is then separated into two regions by the basin boundary, with the asymptotic longtime behaviors approaching  localized turbulent and laminar states. 

Motivated by the experimental findings in bioconvection and similar dynamical structure to the wall-bounded shear flows, 
in this paper, we investigate the onset of the transition in a model system of a dilute suspension of self-propelled particles (SPP), proposed by Childress et. al.\cite{Childress,Taheri} for bioconvection phenomena.
Although the model equations are close to the Rayleigh--B\'{e}nard convection system, a term from the self-motility is added.

In the present study,
we will examine the model bioconvection system, but with 
an extended to a periodic boundary to apply the methodologies based on 
dynamical systems theory. In doing so, we introduce an equilibrium density profile, independently of the self-motility, which enables us to control the effects of
the self-motility in the bulk region.
We will show emergence of the edge state representing localization and bistability
with linear stability analysis.

\section{Setting}
We consider a two-dimensional fluid model with a buoyancy force acting in the $y$ (vertical) direction
defined in a periodic rectangular domain $(x,y) \in [0,8\pi]\times [0,2\pi]$ with an elongation in the $x$ (horizontal) direction.
We consider the Boussinesq approximation for the velocity field, $\bm{u}(x,y,t)$, and 
the continuity equation for the density deviation $m(x,y,t)$ 
from an equilibrium vertical profile $m_0(y)$, we may write down the governing equations in the nondimensionalized form as 
  \begin{align}
    \frac{\partial\bm{u}}{\partial t}+(\bm{u}\cdot\bm{\nabla}) \bm{u}&=\mathrm{Pr}( \nabla^2 \bm{u} -\mathrm{Ra} m\hat{\bm{y}}),  \label{eq:vorticity2}\\
    \frac{\partial m}{\partial t}+(\bm{u}\cdot \bm{\nabla})m&=\nabla^2 m -\mathrm{Pe} \partial_y m -u_y \partial_y m_{0}, \label{eq:density2}
\end{align}
with the incompressibility condition, $\bm{\nabla}\cdot \bm{u}=0$.
Here, the Rayleigh number Ra represents normalized number density, 
and the Prandtl number Pr is the ratio of momentum diffusivity to diffusivity of SPPs. 
Unlike a standard thermal convection system, Eq.\eqref{eq:density2} contains 
an extra term from the self-propelled speed, represented by the P\'{e}clet number Pe, and also the term from the non-uniform density gradient.  

\section{Results}
We now theoretically and numerically investigate the linear stability and nonlinear steady solutions of 
Eqs. \eqref{eq:vorticity2} and \eqref{eq:density2} 
by setting  $n_0 = \sin(y)$ in the two-dimensional periodic box. 
To examine the impact on self-propelled speed represented by Pe,
we change the value of  Pe with fixing Ra (average density), Pr (diffusivity) and size of the domain.

\subsection{Linear stability of trivial solution}

\begin{figure}[th!]
        \begin{center}
          \includegraphics[width=0.6\linewidth]{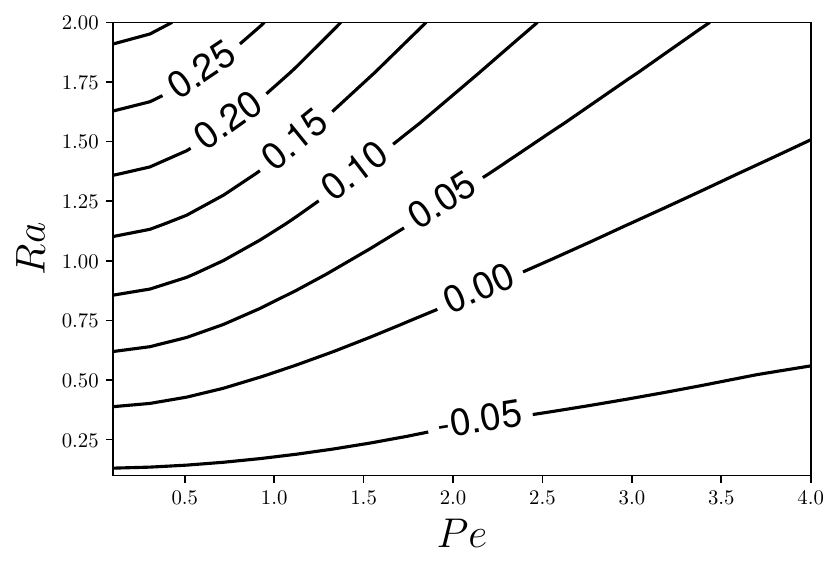}
          \caption{Contour of the maximum growth rate, $\textrm{Re}(\lambda)$, obtained by the linear stability analysis for $Pr =2.5$.
          \label{fig:lin}}
        \end{center}
\end{figure}

We first assume the following form of mode expansion for the vorticity $\omega(x,y,t)$ and the density deviation $m(x,y,t)$:
  \begin{align}
    \omega(x,y,t) &= \exp(\lambda t +ikx ) \sum_{l=-M}^MA_l \exp(ily) \label{eq:exp1},\\
    m(x,y,t) &= \exp(\lambda t +ikx ) \sum_{l=-M}^{M}B_l \exp(ily) \label{eq:exp2},
\end{align}
where $M$ is the cutoff wavenumber for the vertical profile.
By substituting the expansion Eqs. \eqref{eq:exp1} and \eqref{eq:exp2} into Eqs. \eqref{eq:vorticity2} and \eqref{eq:density2},
eigenvalue problem for linear stability is written as
  \begin{align}
    \lambda A_l &= Pr (-k^2 A_l-Ra ik B_l), \label{eq:eig1}\\
    \lambda B_l &= -k^2 B_l-ilPeB_l -\frac{1}{2}\left(\frac{ik A_{l+1}}{k^2+(l+1)^2}+\frac{ik A_{l-1}}{k^2+(l-1)^2}\right) \label{eq:eig2},
\end{align}
for each $l$. We then numerically solved the eigenvalue problem, Eqs. \eqref{eq:eig1} and \eqref{eq:eig2}.
We choose the cut-off wavenumber as $M =32$ and then the linear stability analysis becomes a eigenvalue problem for a square matrix with $2M+1=65$ dimensions.
In Fig.\ref{fig:lin}, we plot the contour of the maximum growth rate, defined as the maximum of the real part of $\lambda$, for the parameter set of $k=0.25$ and $Pr=2.5$. 
The neutral curve is then shown as the contour with $\textrm{Re}(\lambda)=0$.
As seen in Fig.\ref{fig:lin}, all the contour curves increase with Pe, which indicates that the increase of Pe stabilizes the trivial solution.

Besides, we numerically observed that the maximum eigenvalue is always a real value. Additionally, Eqs. \eqref{eq:eig1} and \eqref{eq:eig2} are independent of $Pr$ at $\lambda=0$. By these two observations, the neutral curve should be independent of Pr.

\subsection{Nonlinear solution}

\begin{figure}[th!]
          \includegraphics[width=0.49\linewidth]{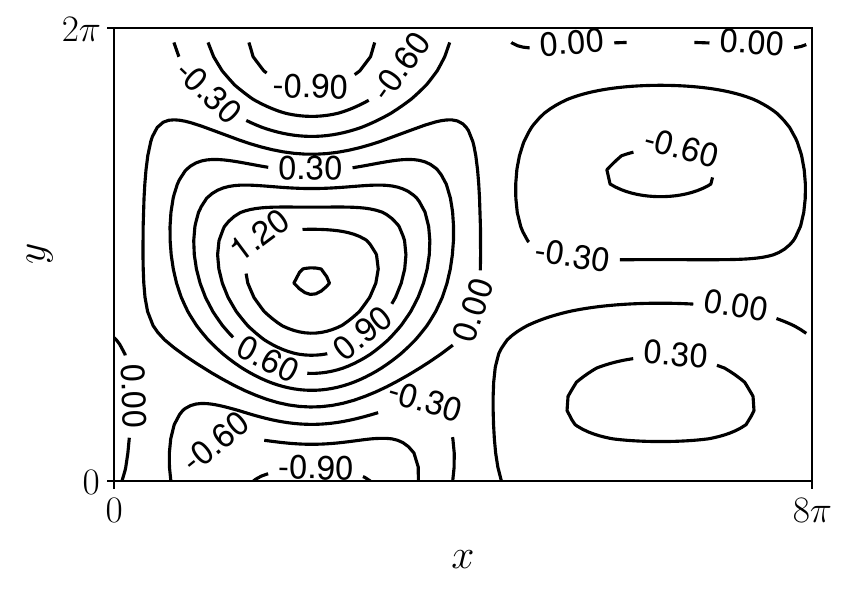}
          \includegraphics[width=0.49\linewidth]{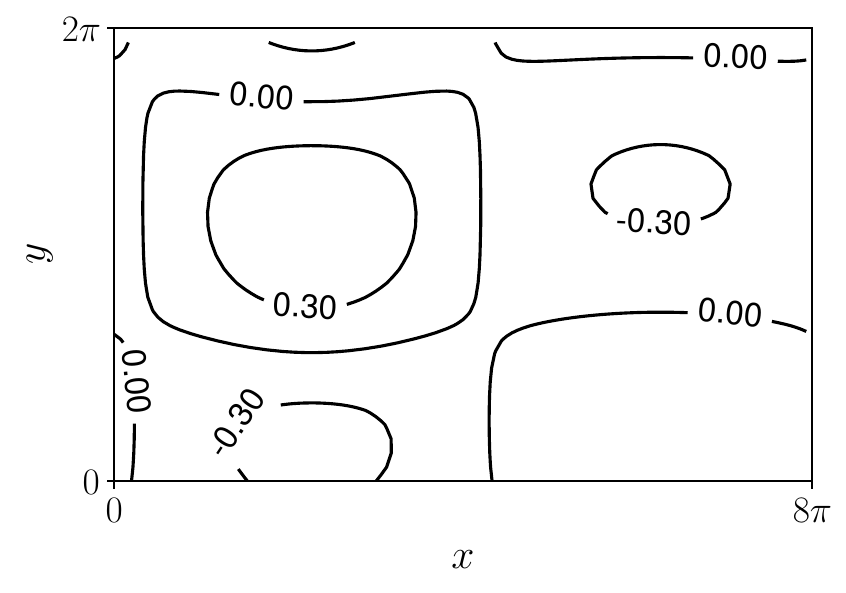}
          \caption{ Density deviation $m(x,y)$ of the nonlinear stationary solutions for $Ra = 0.4$, $Pr =2.5$, and $Pe = 1$. (Left) the stable upper branch solution, $\bm{X}_{UB}$.  (Right) the unstable lower branch solution, $\bm{X}_{LB}$.
           The unstable lower branch solution is obtained by the bisection method.\label{fig:snap}
          }
\end{figure}

We then proceed to examine the system bistability, focusing on the parameter values of $(Ra, Pr, Pe) = (0.4, 2.5, 1)$, where the trivial solution is linearly stable, for an illustrative purpose. 
In this parameter set, we found a stable nonlinear solution [upper branch (UB) solution, $\bm{X}_{UB}$] in addition to the trivial solution.
The density deviation $m(x,y)$ for the UB solution is shown by a  contour map in Fig.~\ref{fig:snap}(left). The density profile is spatially localized, noting the scales are different in the horizontal and vertical directions.

\begin{figure}[th!]
        \begin{center}
          \includegraphics[width=0.6\linewidth]{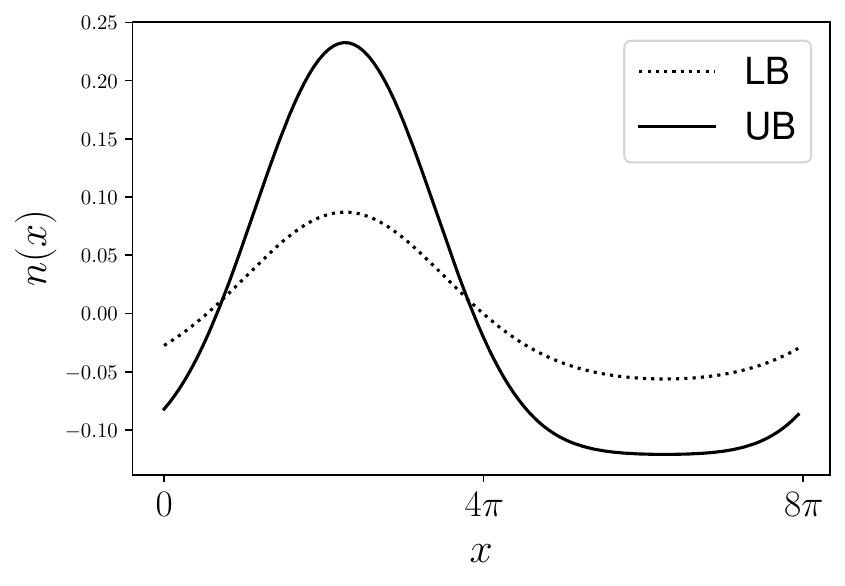}
          \caption{Vertically-averaged density $n(x)$ for the upper and lower branch solutions.\label{fig:ave}}
        \end{center}
\end{figure}

To evaluate the horizontal accumulation, which may be experimentally observed as a localized spot when seen from the top or bottom of a chamber, 
we consider vertically-averaged density, defined as,
    $n(x)=(1/2\pi)\int m(x,y) dy.$
A thick line shown in Fig.\ref{fig:ave} indicates $n(x)$ for the UB solution, which exhibits a 
positive peak at $x\approx 2\pi$, whereas no clear negative peak is observed. 

We then examine the bistability of the system. When two stable solutions coexist in phase space, 
we expect additional unstable solution exists.
In this case, there must exists a basin boundary and an edge state on the basin boundary.

\begin{figure}[t!]
        \begin{center}
          \includegraphics[width=0.6\linewidth]{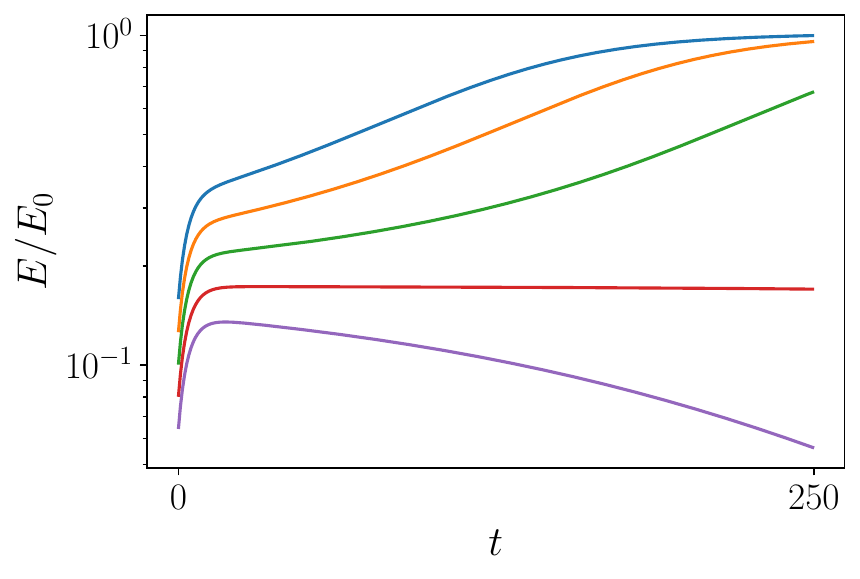}
          \caption{ Bisection method to obtain a state on basin boundary. Time evolution of energy norm of the velocity, normalized by the energy norm of the UB solution, $E_0$, for $Ra = 0.4$, $Pr =2.5$, and $Pe = 1$. The different colors corresponds to different initial conditions, $\bm{X}_\alpha$. The red curve with stationary energy norm is the unstable lower branch (LB) solution, $\bm{X}_{LB}$, which is identified as an edge state. \label{fig:bisec}}
        \end{center}
\end{figure}

To obtain an unstable solution in a basin boundary,
we perform the bisection iteration method used in the previous study of wall-bounded flow \cite{Itano}, and introduce $\bm{X}_{\alpha}=\alpha \bm{X}_{UB}$ with an additional parameter $\alpha\in [0, 1]$ and use this as our initial conditions.
Since $\bm{X}_{0}$ is the trivial solution, there exists $\alpha_c\in[0,1]$ such that $\bm{X}_{\alpha_c}$ is located on the basin boundary.
In the bisection method, we seek a state with the energy norm 
$E=(u_x^2+u_y^2)/2$ of $\bm{X}_{\alpha_c}$ being unchanged for a long time, and  such $\alpha_c$ has been numerically detected [Fig. \ref{fig:bisec}].
As seen in Fig.\ref{fig:bisec}, at the particular value of  $\alpha=\alpha_c$, the energy norm $E$ is unchanged after initial transient time duration, as shown in the red line.
Once $\alpha$ exceeds the critical value, $\alpha>\alpha_c$, the energy norm $E$ approaches $E_0$, the energy norm of the UB solution, i.e.,
$X_{UB}$ is an attracting state. If $\alpha<\alpha_c$, on the other hand, $E$ decays to zero,
being consistent with the stable trivial solution.
The state $\bm{X}_{\alpha_c}$ is found to be an unstable stationary solution on the basin boundary, hence, an edge state.
We refer this solution as the lower blanch (LB) $\bm{X}_{LB}$ solution hereafter.

The density deviation of the LB solution for the parameter set, $(Ra, Pr, Pe) = (0.4, 2.5, 1)$, is shown in Fig.~\ref{fig:snap}(Right).
As seen in the UB solution [Fig.~\ref{fig:snap}(Left)], the LB solution is also spatially localized.
However, the absolute values of $m(x,y)$ are relatively smaller than those of LB solution.
The dotted line shown in Fig.\ref{fig:ave} indicates vertically-averaged density $n(x)$ for the LB solution, which exhibits a weakly localized peak compared with the UB solution.

\section{Conclusions}
In this study, to understand the effect of the the self-propelled speed on emergence of localized, bistable flow patterns of a suspension of SPPs,
we introduced a model system motivated by bioconvection in a doubly periodic box with a
fixed equilibrium density profile in the vertical direction.
Linear stability analysis of the model convection system clarifies that an increase of self-motility in the vertical direction results in 
stabilizing the trivial nonconvection state.
Even when the trivial solution is linearly stable, nonlinear solution 
can be stable, unlike Rayleigh--B\'{e}nard convection.
This localized, bistable nature is analogous to the turbulence transition in shear flows. By using the methodology based on the dynamical systems theory developed in these systems, we numerically obtained an unstable stationary solution which is identified as an edge state, whose stable manifold forms a basin boundary between two stable solutions.\\

{\bf Acknowledgments}\\
K.I. acknowledges the Japan Society for the Promotion of Science (JSPS) KAKENHI (Grant No. 21H05309) and the Japan Science and Technology Agency (JST), FOREST (Grant No. JPMJFR212N).

%
%

\end{document}